\documentclass[a4paper,11pt]{article}
\usepackage{pos}

\usepackage[symbol]{footmisc}
\usepackage[english]{babel}
\usepackage{graphicx}
\usepackage{graphics}
\usepackage{braket}
\usepackage{bbold}
\usepackage{amsmath}
\usepackage{nicefrac}
\usepackage{dcolumn}
\usepackage{bm}
\usepackage{slashed}
\usepackage{datetime}
\usepackage{mciteplus}
\usepackage{multirow}
\usepackage{bbold}
\usepackage{siunitx}
\usepackage{booktabs}
\usepackage{color, soul}
\usepackage[usenames,dvipsnames]{xcolor}
\usepackage{float}
\usepackage[utf8]{inputenc}
\usepackage[normalem]{ulem}
\usepackage{mathtools}
\usepackage{setspace}

\renewcommand{\thefootnote}{\fnsymbol{footnote}}

\newcommand{\LQCD}{\Lambda_{\rm QCD}}

\newcommand{\DY}{\Delta Y}

\title{Higgs production at NLL accuracy in the BFKL approach}
\ShortTitle{Higgs production at NLL accuracy in the BFKL approach}

\author*[a]{Francesco Giovanni Celiberto}
\author[b,c]{Luigi Delle Rose}
\author[d]{Michael Fucilla}
\author[b,c]{\\Gabriele Gatto}
\author[e]{Dmitry Yu. Ivanov}
\author[b,c]{Mohammed M. A. Mohammed}
\author[b,c]{Alessandro Papa}

\affiliation[a]{Universidad de Alcalá (UAH), Departamento de Física y Matemáticas, Campus Universitario, \\ Alcalá de Henares, E-28805, Madrid, Spain}

\affiliation[b]{Dipartimento di Fisica, Università della Calabria, Arcavacata di Rende, I-87036, Cosenza, Italy}

\affiliation[c]{
INFN, Gruppo Collegato di Cosenza, Arcavacata di Rende, I-87036, Cosenza, Italy}

\affiliation[d]{Université Paris-Saclay, CNRS/IN2P3, IJCLab, 91405, Orsay, France}

\affiliation[e]{Sobolev Institute of Mathematics, 630090 Novosibirsk, Russia}

\emailAdd{francesco.celiberto@uah.es}
\emailAdd{luigi.dellerose@unical.it}
\emailAdd{michael.fucilla@ijclab.in2p3.fr}
\emailAdd{gabriele.gatto@unical.it}
\emailAdd{d-ivanov@math.nsc.ru}
\emailAdd{mohammed.maher@unical.it}
\emailAdd{alessandro.papa@fis.unical.it}

\abstract{Precision physics in the Higgs sector has been one of the main challenges of particle physics in the recent years. The pure fixed-order calculations entering the collinear factorization framework, which have been pushed up to next-cube-leading-order, are not able to describe the entire kinematic spectrum. In particular sectors, they have to be necessarily enhanced by all-order resummations. 
In the so-called semi-hard regime, large energy-type logarithms spoil the perturbative convergence of the series and must be resummed to all orders. 
This resummation is a core ingredient for a correct description of the inclusive hadroproduction of a forward Higgs boson in the limit of small Bjorken $x$, as well as for a precision study of inclusive forward emissions of a Higgs boson in association with a backward identified object. A complete resummation for these processes can be achieved at the at next-to-leading logarithmic accuracy thanks to the Balitsky--Fadin--Kuraev--Lipatov approach.
In the present work we present and discuss a series of recent phenomenological results within a partial next-to-leading accuracy. 
They include the analysis of rapidity and azimuthal-angle differential rates for Higgs plus jet and Higgs plus charm reactions in forward and ultraforward directions of rapidity at the LHC.}

\FullConference{31st International Workshop on Deep Inelastic Scattering (DIS2024)\\
 8–12 April 2024\\
Grenoble, France}

\begin{document}
\maketitle

\section{Hors d'{\oe}uvre}
\label{sec:introduction}

The \emph{all-order} resummation of high-energy logarithms represents a valuable tool for a precise description of semi-inclusive Higgs-production rates at the LHC as well as the future FCC.
In the \emph{semi-hard} regime of QCD, the stringent scale order, $\sqrt{s} \gg \{\mu_i\} \gg \LQCD$, with $\{\mu_i\}$ being a set of process-characteristic hard scales and $\sqrt{s}$ standing for the center-of-mass energy, heightens the weight of energy logarithms.
The Balitsky--Fadin--Kuraev--Lipatov (BFKL) approach~\cite{Fadin:1975cb,Kuraev:1976ge,Balitsky:1978ic} offers us a powerful way to resum those logarithms at the leading-logarithmic (LL) and next-to-leading logarithmic (NLL) level.
It also allows us to probe the gluon content of the proton at low $x$~\cite{Bacchetta:2020vty,Bacchetta:2024fci,Bacchetta:2021lvw,Bacchetta:2021twk,Bacchetta:2024uxb,Arbuzov:2020cqg,Celiberto:2021zww,Amoroso:2022eow,Bolognino:2018rhb,Bolognino:2021niq,Hentschinski:2022xnd,Celiberto:2019slj,Bolognino:2022uty}.
Semi-inclusive hadroproductions of two particles tagged with large transverse masses and a high rapidity separation, $\DY$, stand as a promising testing ground of high-energy QCD.
To study these two-particle processes, a \emph{multilateral} factorization, where both high-energy and collinear dynamics come into play, is needed.
To this scope, a \emph{hybrid} factorization formalism (HyF) was developed~\cite{Celiberto:2020tmb,Bolognino:2021mrc} (see also~\cite{vanHameren:2022mtk,Bonvini:2018ixe,Silvetti:2022hyc} for single-particle detections).
HyF cross sections feature a transverse-momentum convolution of the universal BFKL Green's function with two process-related impact factors.
The latters read in turn as a sub-convolution of singly off-shell coefficient functions and collinear parton distributions (PDFs).
Phenomenological studies of the HyF formalism within a full or partial NLL accuracy were done through: Mueller--Navelet jet emissions~\cite{Ducloue:2013hia,Ducloue:2013bva,Celiberto:2015yba,Celiberto:2015mpa,Celiberto:2016ygs,Celiberto:2017ius,Caporale:2018qnm,Celiberto:2022gji,Egorov:2023duz,Baldenegro:2024ndr}, Drell--Yan pair~\cite{Motyka:2014lya,Motyka:2016lta,Celiberto:2018muu,Golec-Biernat:2018kem}, light~\cite{Celiberto:2016hae,Celiberto:2017ptm,Bolognino:2018oth,Bolognino:2019yqj,Bolognino:2019cac,Celiberto:2020rxb,Celiberto:2020wpk,Celiberto:2022kxx} or heavy-light~\cite{Celiberto:2017nyx,Bolognino:2019yls,Bolognino:2019ccd,AlexanderAryshev:2022pkx,Celiberto:2021dzy,Celiberto:2021fdp,Celiberto:2022zdg,Celiberto:2022keu,Celiberto:2024omj,Anchordoqui:2021ghd,Feng:2022inv,Celiberto:2022rfj,Celiberto:2024swu} hadron, quarkonium~\cite{Boussarie:2017oae,Chapon:2020heu,Celiberto:2022dyf,Celiberto:2023fzz}, and exotic-matter~\cite{Celiberto:2023rzw,Celiberto:2024mab,Celiberto:2024mrq} detections.
In this work we will study the semi-inclusive tag of a forward Higgs boson accompanied by a light-flavored jet~\cite{Celiberto:2020tmb} (for corresponding next-to-next-to-leading analyses without resummations, or next-to-NLL  investigations within the transverse-momentum resummation formalism, see~\cite{Chen:2014gva,Boughezal:2015dra} and~\cite{Monni:2019yyr}, respectively).
We will go with a partial NLL accuracy, which relies upon the NLL Green's function plus leading-order coefficient functions.

\section{Higgs production at NLL accuracy}
\label{sec:results}

\begin{figure*}[!t]
\centering

\includegraphics[scale=0.465,clip]{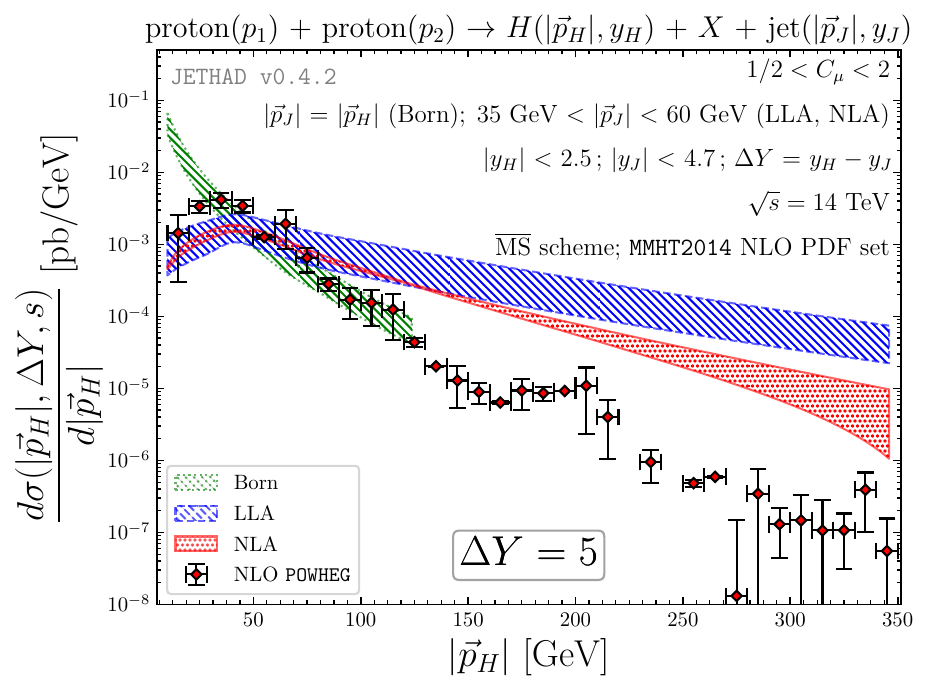}
 \hspace{0.10cm}
\includegraphics[scale=0.465,clip]{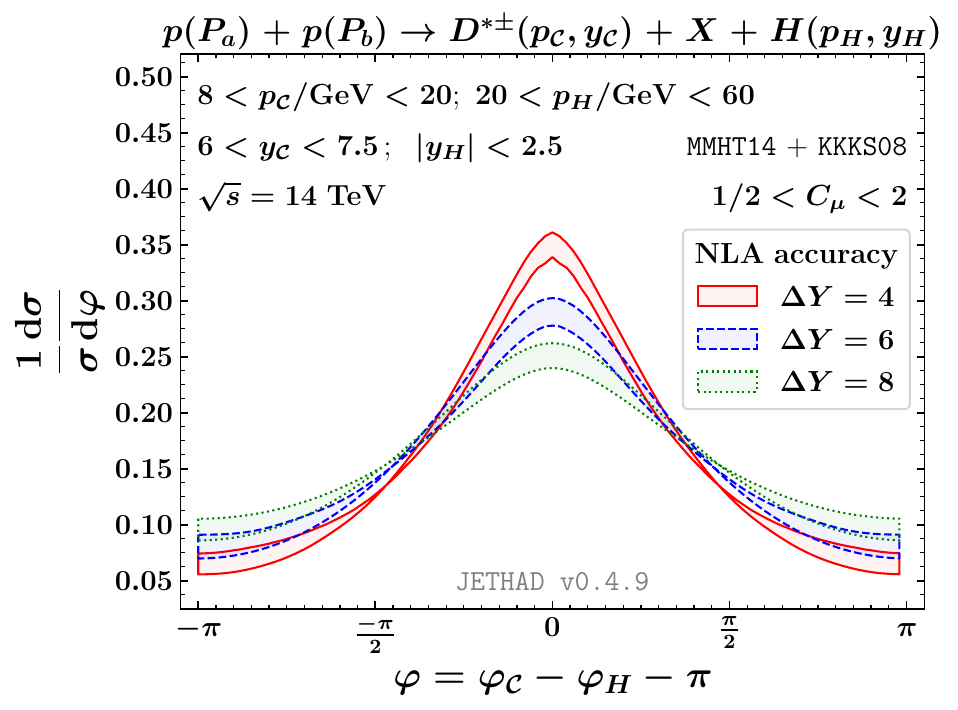}

\caption{Left panel: Higgs plus jet transverse-momentum spectrum at 14~TeV LHC.
Right panel: Higgs plus $D^{*\pm}$ meson angular multiplicity at 14~TeV FPF~$+$~ATLAS.
Uncertainty bands show $\mu_{R,F}$ variation in the $1 < C_{\mu} < 2$ range. Text boxes refer to kinematic cuts.}

\label{fig:pheno}
\end{figure*}

Left panel of Fig.~\ref{fig:pheno} shows the Higgs plus jet hadroproduction rate at 14~TeV, differential in the transverse momentum of the Higgs boson, $|\vec p_H|$, and taken at $\DY = 5$.
Rapidity ranges are the typical one of CMS or ATLAS studies, with the Higgs boson detected only in the barrel calorimeter ($|y_H| < 2.5$) and the jet reconstructed also by the endcaps ($|y_J| < 4.7$).
We observe that, in the BFKL-expected kinematic sector, namely the peak region plus the first part of the distribution tail, where $|\vec p_H| \sim |\vec p_J|$, resummed predictions are quite stable under energy scale variations, with NLL uncertainty bands (red) almost completely contained inside pure LL ones (blue).
This brings clear evidence that the emission of a Higgs boson acts as a \emph{natural stabilizer} of the high-energy resummation~\cite{Celiberto:2020tmb,Celiberto:2021tky,Celiberto:2021fjf,Celiberto:2021txb}.
Conversely, in the large $|\vec p_H|$-tail, NLL BFKL decouples from its LL limit and the corresponding uncertainty band becomes wider and wider with $|\vec p_H|$.
This happens because, in this kinematic sector, large DGLAP-type as well as \emph{threshold} logarithms, not accounted for by our formalisms, are enhanced.
We also note that NLL results are qualitatively close to NLO fixed-order ones from POWHEG~\cite{Hamilton:2012rf,Bagnaschi:2023rbx,Banfi:2023mhz} only in the peak region.
This is a clear signal that, to get a precise description of our high-energy observables, a \emph{matching} between the NLL HyF formalism and the NLO background is needed~\cite{Celiberto:2023uuk_article,Celiberto:2023eba_article,Celiberto:2023nym}.
Right panel of Fig.~\ref{fig:pheno} shows the Higgs plus $D^{*\pm}$ NLL azimuthal multiplicity at 14~TeV, for different values of $\DY$ and with the $D^{*\pm}$ meson detected in the ultraforward rapidity directions ($6 < y_{\cal C} < 7.5$) reachable at the planned Forward Physics Facility~\cite{Anchordoqui:2021ghd,Feng:2022inv,Celiberto:2022zdg}.
We note that, as $\DY$ grows, distribution peaks shrink while their widths moderately widen.
This is a clear signal of the onset of BFKL dynamics.
Indeed, larger and larger values of $\DY$ heighten the weight of secondary gluons strongly ordered in rapidity, whose effect is caught by the BFKL resummation.

\section{Closing statements}
\label{sec:conclusions}

We have studied the production of a Higgs boson, accompanied by a jet~\cite{Celiberto:2020tmb} or a singly charmed hadron~\cite{Celiberto:2022zdg} in (ultra)forward directions of rapidity at 14~TeV LHC.
Future analyses will include: $(i)$ NLO contributions to the Higgs impact factor~\cite{Hentschinski:2020tbi,Celiberto:2022fgx,Nefedov:2019mrg}, $(ii)$ a \emph{matching} with the fixed-order signal~\cite{Celiberto:2023uuk_article,Celiberto:2023eba_article,Celiberto:2023nym}, and $(iii)$ a phenomenological extension to nominal FCC energies~\cite{Dawson:2022zbb,Celiberto:2023rtu}.



\begingroup
\setstretch{0.6}
\bibliographystyle{bibstyle}
\bibliography{biblography}

\begin{thebibliography}{80}
\expandafter\ifx\csname natexlab\endcsname\relax\def\natexlab#1{#1}\fi
\expandafter\ifx\csname bibnamefont\endcsname\relax
  \def\bibnamefont#1{#1}\fi
\expandafter\ifx\csname bibfnamefont\endcsname\relax
  \def\bibfnamefont#1{#1}\fi
\expandafter\ifx\csname citenamefont\endcsname\relax
  \def\citenamefont#1{#1}\fi
\expandafter\ifx\csname url\endcsname\relax
  \def\url#1{\texttt{#1}}\fi
\expandafter\ifx\csname urlprefix\endcsname\relax\def\urlprefix{URL }\fi
\providecommand{\bibinfo}[2]{#2}
\providecommand{\eprint}[2][]{\url{#2}}

\bibitem[{\citenamefont{Fadin et~al.}(1975)}]{Fadin:1975cb}
\bibinfo{author}{\bibfnamefont{V.~S.} \bibnamefont{Fadin}} \bibnamefont{et~al.}, \bibinfo{journal}{Phys. Lett. B} \textbf{\bibinfo{volume}{60}}, \bibinfo{pages}{50} (\bibinfo{year}{1975}).

\bibitem[{\citenamefont{Kuraev et~al.}(1976)}]{Kuraev:1976ge}
\bibinfo{author}{\bibfnamefont{E.~A.} \bibnamefont{Kuraev}} \bibnamefont{et~al.}, \bibinfo{journal}{Sov. Phys. JETP} \textbf{\bibinfo{volume}{44}}, \bibinfo{pages}{443} (\bibinfo{year}{1976}).

\bibitem[{\citenamefont{Balitsky\mbox{, L. N. Lipatov}}(1978)}]{Balitsky:1978ic}
\bibinfo{author}{\bibfnamefont{I.~I.} \bibnamefont{Balitsky\mbox{, L. N. Lipatov}}}, \bibinfo{journal}{Sov.\ J.\ Nucl.\ Phys.} \textbf{\bibinfo{volume}{28}}, \bibinfo{pages}{822} (\bibinfo{year}{1978}).

\bibitem[{\citenamefont{Bacchetta et~al.}(2020)}]{Bacchetta:2020vty}
\bibinfo{author}{\bibfnamefont{A.}~\bibnamefont{Bacchetta}} \bibnamefont{et~al.}, \bibinfo{journal}{Eur. Phys. J. C} \textbf{\bibinfo{volume}{80}}, \bibinfo{pages}{733} (\bibinfo{year}{2020}), \eprint{2005.02288}.

\bibitem[{\citenamefont{Bacchetta et~al.}(2024{\natexlab{a}})\citenamefont{Bacchetta, Celiberto,  Radici}}]{Bacchetta:2024fci}
\bibinfo{author}{\bibfnamefont{A.}~\bibnamefont{Bacchetta}}, \bibinfo{author}{\bibfnamefont{F.~G.} \bibnamefont{Celiberto}},  \bibinfo{author}{\bibfnamefont{M.}~\bibnamefont{Radici}}, \bibinfo{journal}{Eur. Phys. J. C} \textbf{\bibinfo{volume}{84}}, \bibinfo{pages}{576} (\bibinfo{year}{2024}{\natexlab{a}}), \eprint{2402.17556}.

\bibitem[{\citenamefont{Bacchetta et~al.}(2022{\natexlab{a}})\citenamefont{Bacchetta, Celiberto,  Radici}}]{Bacchetta:2021lvw}
\bibinfo{author}{\bibfnamefont{A.}~\bibnamefont{Bacchetta}}, \bibinfo{author}{\bibfnamefont{F.~G.} \bibnamefont{Celiberto}},  \bibinfo{author}{\bibfnamefont{M.}~\bibnamefont{Radici}}, \bibinfo{journal}{PoS} \textbf{\bibinfo{volume}{EPS-HEP2021}}, \bibinfo{pages}{376} (\bibinfo{year}{2022}{\natexlab{a}}), \eprint{2111.01686}.

\bibitem[{\citenamefont{Bacchetta et~al.}(2022{\natexlab{b}})\citenamefont{Bacchetta, Celiberto,  Radici}}]{Bacchetta:2021twk}
\bibinfo{author}{\bibfnamefont{A.}~\bibnamefont{Bacchetta}}, \bibinfo{author}{\bibfnamefont{F.~G.} \bibnamefont{Celiberto}},  \bibinfo{author}{\bibfnamefont{M.}~\bibnamefont{Radici}}, \bibinfo{journal}{PoS} \textbf{\bibinfo{volume}{PANIC2021}}, \bibinfo{pages}{378} (\bibinfo{year}{2022}{\natexlab{b}}), \eprint{2111.03567}.

\bibitem[{\citenamefont{Bacchetta et~al.}(2024{\natexlab{b}})\citenamefont{Bacchetta, Celiberto,  Radici}}]{Bacchetta:2024uxb}
\bibinfo{author}{\bibfnamefont{A.}~\bibnamefont{Bacchetta}}, \bibinfo{author}{\bibfnamefont{F.~G.} \bibnamefont{Celiberto}},  \bibinfo{author}{\bibfnamefont{M.}~\bibnamefont{Radici}}, \bibinfo{journal}{PoS} \textbf{\bibinfo{volume}{SPIN2023}}, \bibinfo{pages}{049} (\bibinfo{year}{2024}{\natexlab{b}}), \eprint{2406.04893}.

\bibitem[{\citenamefont{Arbuzov et~al.}(2021)}]{Arbuzov:2020cqg}
\bibinfo{author}{\bibfnamefont{A.}~\bibnamefont{Arbuzov}} \bibnamefont{et~al.}, \bibinfo{journal}{Prog. Part. Nucl. Phys.} \textbf{\bibinfo{volume}{119}}, \bibinfo{pages}{103858} (\bibinfo{year}{2021}), \eprint{2011.15005}.

\bibitem[{\citenamefont{Celiberto}(2021{\natexlab{a}})}]{Celiberto:2021zww}
\bibinfo{author}{\bibfnamefont{F.~G.} \bibnamefont{Celiberto}}, \bibinfo{journal}{Nuovo Cim.} \textbf{\bibinfo{volume}{C44}}, \bibinfo{pages}{36} (\bibinfo{year}{2021}{\natexlab{a}}), \eprint{2101.04630}.

\bibitem[{\citenamefont{Amoroso et~al.}(2022)}]{Amoroso:2022eow}
\bibinfo{author}{\bibfnamefont{S.}~\bibnamefont{Amoroso}} \bibnamefont{et~al.}, \bibinfo{journal}{Acta Phys. Polon. B} \textbf{\bibinfo{volume}{53}}, \bibinfo{pages}{A1} (\bibinfo{year}{2022}), \eprint{2203.13923}.

\bibitem[{\citenamefont{Bolognino et~al.}(2018{\natexlab{a}})}]{Bolognino:2018rhb}
\bibinfo{author}{\bibfnamefont{A.~D.} \bibnamefont{Bolognino}} \bibnamefont{et~al.}, \bibinfo{journal}{Eur. Phys. J.} \textbf{\bibinfo{volume}{C78}}, \bibinfo{pages}{1023} (\bibinfo{year}{2018}{\natexlab{a}}), \eprint{1808.02395}.

\bibitem[{\citenamefont{Bolognino et~al.}(2021{\natexlab{a}})}]{Bolognino:2021niq}
\bibinfo{author}{\bibfnamefont{A.~D.} \bibnamefont{Bolognino}} \bibnamefont{et~al.}, \bibinfo{journal}{Eur. Phys. J. C} \textbf{\bibinfo{volume}{81}}, \bibinfo{pages}{846} (\bibinfo{year}{2021}{\natexlab{a}}), \eprint{2107.13415}.

\bibitem[{\citenamefont{Hentschinski et~al.}(2023)}]{Hentschinski:2022xnd}
\bibinfo{author}{\bibfnamefont{M.}~\bibnamefont{Hentschinski}} \bibnamefont{et~al.}, \bibinfo{journal}{Acta Phys. Polon. B} \textbf{\bibinfo{volume}{54}}, \bibinfo{pages}{2} (\bibinfo{year}{2023}), \eprint{2203.08129}.

\bibitem[{\citenamefont{Celiberto}(2019)}]{Celiberto:2019slj}
\bibinfo{author}{\bibfnamefont{F.~G.} \bibnamefont{Celiberto}}, \bibinfo{journal}{Nuovo Cim.} \textbf{\bibinfo{volume}{C42}}, \bibinfo{pages}{220} (\bibinfo{year}{2019}), \eprint{1912.11313}.

\bibitem[{\citenamefont{Bolognino et~al.}(2022)}]{Bolognino:2022uty}
\bibinfo{author}{\bibfnamefont{A.~D.} \bibnamefont{Bolognino}} \bibnamefont{et~al.}, \bibinfo{journal}{Rev. Mex. Fis. Suppl.} \textbf{\bibinfo{volume}{3}}, \bibinfo{pages}{0308109} (\bibinfo{year}{2022}), \eprint{2202.02513}.

\bibitem[{\citenamefont{Celiberto et~al.}(2021{\natexlab{a}})}]{Celiberto:2020tmb}
\bibinfo{author}{\bibfnamefont{F.~G.} \bibnamefont{Celiberto}} \bibnamefont{et~al.}, \bibinfo{journal}{Eur. Phys. J. C} \textbf{\bibinfo{volume}{81}}, \bibinfo{pages}{293} (\bibinfo{year}{2021}{\natexlab{a}}), \eprint{2008.00501}.

\bibitem[{\citenamefont{Bolognino et~al.}(2021{\natexlab{b}})}]{Bolognino:2021mrc}
\bibinfo{author}{\bibfnamefont{A.~D.} \bibnamefont{Bolognino}} \bibnamefont{et~al.}, \bibinfo{journal}{Phys. Rev. D} \textbf{\bibinfo{volume}{103}}, \bibinfo{pages}{094004} (\bibinfo{year}{2021}{\natexlab{b}}), \eprint{2103.07396}.

\bibitem[{\citenamefont{van Hameren et~al.}(2022)\citenamefont{van Hameren, Motyka,  Ziarko}}]{vanHameren:2022mtk}
\bibinfo{author}{\bibfnamefont{A.}~\bibnamefont{van Hameren}}, \bibinfo{author}{\bibfnamefont{L.}~\bibnamefont{Motyka}},  \bibinfo{author}{\bibfnamefont{G.}~\bibnamefont{Ziarko}}, \bibinfo{journal}{JHEP} \textbf{\bibinfo{volume}{11}}, \bibinfo{pages}{103} (\bibinfo{year}{2022}), \eprint{2205.09585}.

\bibitem[{\citenamefont{Bonvini\mbox{, S. Marzani}}(2018)}]{Bonvini:2018ixe}
\bibinfo{author}{\bibfnamefont{M.}~\bibnamefont{Bonvini\mbox{, S. Marzani}}}, \bibinfo{journal}{Phys. Rev. Lett.} \textbf{\bibinfo{volume}{120}}, \bibinfo{pages}{202003} (\bibinfo{year}{2018}), \eprint{1802.07758}.

\bibitem[{\citenamefont{Silvetti\mbox{, M. Bonvini}}(2023)}]{Silvetti:2022hyc}
\bibinfo{author}{\bibfnamefont{F.}~\bibnamefont{Silvetti\mbox{, M. Bonvini}}}, \bibinfo{journal}{Eur. Phys. J. C} \textbf{\bibinfo{volume}{83}}, \bibinfo{pages}{267} (\bibinfo{year}{2023}), \eprint{2211.10142}.

\bibitem[{\citenamefont{Duclou\'e et~al.}(2013)\citenamefont{Duclou\'e, Szymanowski,  Wallon}}]{Ducloue:2013hia}
\bibinfo{author}{\bibfnamefont{B.}~\bibnamefont{Duclou\'e}}, \bibinfo{author}{\bibfnamefont{L.}~\bibnamefont{Szymanowski}},  \bibinfo{author}{\bibfnamefont{S.}~\bibnamefont{Wallon}}, \bibinfo{journal}{JHEP} \textbf{\bibinfo{volume}{05}}, \bibinfo{pages}{096} (\bibinfo{year}{2013}), \eprint{1302.7012}.

\bibitem[{\citenamefont{Duclou\'e et~al.}(2014)\citenamefont{Duclou\'e, Szymanowski,  Wallon}}]{Ducloue:2013bva}
\bibinfo{author}{\bibfnamefont{B.}~\bibnamefont{Duclou\'e}}, \bibinfo{author}{\bibfnamefont{L.}~\bibnamefont{Szymanowski}},  \bibinfo{author}{\bibfnamefont{S.}~\bibnamefont{Wallon}}, \bibinfo{journal}{Phys. Rev. Lett.} \textbf{\bibinfo{volume}{112}}, \bibinfo{pages}{082003} (\bibinfo{year}{2014}), \eprint{1309.3229}.

\bibitem[{\citenamefont{Celiberto et~al.}(2015{\natexlab{a}})}]{Celiberto:2015yba}
\bibinfo{author}{\bibfnamefont{F.~G.} \bibnamefont{Celiberto}} \bibnamefont{et~al.}, \bibinfo{journal}{Eur. Phys. J. C} \textbf{\bibinfo{volume}{75}}, \bibinfo{pages}{292} (\bibinfo{year}{2015}{\natexlab{a}}), \eprint{1504.08233}.

\bibitem[{\citenamefont{Celiberto et~al.}(2015{\natexlab{b}})}]{Celiberto:2015mpa}
\bibinfo{author}{\bibfnamefont{F.~G.} \bibnamefont{Celiberto}} \bibnamefont{et~al.}, \bibinfo{journal}{Acta Phys. Polon. Supp.} \textbf{\bibinfo{volume}{8}}, \bibinfo{pages}{935} (\bibinfo{year}{2015}{\natexlab{b}}), \eprint{1510.01626}.

\bibitem[{\citenamefont{Celiberto et~al.}(2016{\natexlab{a}})}]{Celiberto:2016ygs}
\bibinfo{author}{\bibfnamefont{F.~G.} \bibnamefont{Celiberto}} \bibnamefont{et~al.}, \bibinfo{journal}{Eur. Phys. J. C} \textbf{\bibinfo{volume}{76}}, \bibinfo{pages}{224} (\bibinfo{year}{2016}{\natexlab{a}}), \eprint{1601.07847}.

\bibitem[{\citenamefont{Celiberto}(2017)}]{Celiberto:2017ius}
\bibinfo{author}{\bibfnamefont{F.~G.} \bibnamefont{Celiberto}}, Ph.D. thesis (\bibinfo{year}{2017}), \eprint{1707.04315}.

\bibitem[{\citenamefont{Caporale et~al.}(2018)}]{Caporale:2018qnm}
\bibinfo{author}{\bibfnamefont{F.}~\bibnamefont{Caporale}} \bibnamefont{et~al.}, \bibinfo{journal}{Nucl. Phys. B} \textbf{\bibinfo{volume}{935}}, \bibinfo{pages}{412} (\bibinfo{year}{2018}), \eprint{1806.06309}.

\bibitem[{\citenamefont{Celiberto{, A. Papa}}(2022)}]{Celiberto:2022gji}
\bibinfo{author}{\bibfnamefont{F.~G.} \bibnamefont{Celiberto{, A. Papa}}}, \bibinfo{journal}{Phys. Rev. D} \textbf{\bibinfo{volume}{106}}, \bibinfo{pages}{114004} (\bibinfo{year}{2022}), \eprint{2207.05015}.

\bibitem[{\citenamefont{Egorov{, V. T. Kim}}(2023)}]{Egorov:2023duz}
\bibinfo{author}{\bibfnamefont{A.~I.} \bibnamefont{Egorov{, V. T. Kim}}}, \bibinfo{journal}{Phys. Rev. D} \textbf{\bibinfo{volume}{108}}, \bibinfo{pages}{014010} (\bibinfo{year}{2023}), \eprint{2305.19854}.

\bibitem[{\citenamefont{Baldenegro et~al.}(2024)}]{Baldenegro:2024ndr}
\bibinfo{author}{\bibfnamefont{C.}~\bibnamefont{Baldenegro}} \bibnamefont{et~al.} (\bibinfo{year}{2024}), \eprint{2406.10681}.

\bibitem[{\citenamefont{Motyka et~al.}(2015)\citenamefont{Motyka, Sadzikowski,  Stebel}}]{Motyka:2014lya}
\bibinfo{author}{\bibfnamefont{L.}~\bibnamefont{Motyka}}, \bibinfo{author}{\bibfnamefont{M.}~\bibnamefont{Sadzikowski}},  \bibinfo{author}{\bibfnamefont{T.}~\bibnamefont{Stebel}}, \bibinfo{journal}{JHEP} \textbf{\bibinfo{volume}{05}}, \bibinfo{pages}{087} (\bibinfo{year}{2015}), \eprint{1412.4675}.

\bibitem[{\citenamefont{Motyka et~al.}(2017)\citenamefont{Motyka, Sadzikowski,  Stebel}}]{Motyka:2016lta}
\bibinfo{author}{\bibfnamefont{L.}~\bibnamefont{Motyka}}, \bibinfo{author}{\bibfnamefont{M.}~\bibnamefont{Sadzikowski}},  \bibinfo{author}{\bibfnamefont{T.}~\bibnamefont{Stebel}}, \bibinfo{journal}{Phys. Rev.} \textbf{\bibinfo{volume}{D95}}, \bibinfo{pages}{114025} (\bibinfo{year}{2017}), \eprint{1609.04300}.

\bibitem[{\citenamefont{Celiberto et~al.}(2018{\natexlab{a}})}]{Celiberto:2018muu}
\bibinfo{author}{\bibfnamefont{F.~G.} \bibnamefont{Celiberto}} \bibnamefont{et~al.}, \bibinfo{journal}{Phys. Lett.} \textbf{\bibinfo{volume}{B786}}, \bibinfo{pages}{201} (\bibinfo{year}{2018}{\natexlab{a}}), \eprint{1808.09511}.

\bibitem[{\citenamefont{Golec-Biernat et~al.}(2018)}]{Golec-Biernat:2018kem}
\bibinfo{author}{\bibfnamefont{K.}~\bibnamefont{Golec-Biernat}} \bibnamefont{et~al.}, \bibinfo{journal}{JHEP} \textbf{\bibinfo{volume}{12}}, \bibinfo{pages}{091} (\bibinfo{year}{2018}), \eprint{1811.04361}.

\bibitem[{\citenamefont{Celiberto et~al.}(2016{\natexlab{b}})}]{Celiberto:2016hae}
\bibinfo{author}{\bibfnamefont{F.~G.} \bibnamefont{Celiberto}} \bibnamefont{et~al.}, \bibinfo{journal}{Phys. Rev. D} \textbf{\bibinfo{volume}{94}}, \bibinfo{pages}{034013} (\bibinfo{year}{2016}{\natexlab{b}}), \eprint{1604.08013}.

\bibitem[{\citenamefont{Celiberto et~al.}(2017)}]{Celiberto:2017ptm}
\bibinfo{author}{\bibfnamefont{F.~G.} \bibnamefont{Celiberto}} \bibnamefont{et~al.}, \bibinfo{journal}{Eur. Phys. J. C} \textbf{\bibinfo{volume}{77}}, \bibinfo{pages}{382} (\bibinfo{year}{2017}), \eprint{1701.05077}.

\bibitem[{\citenamefont{Bolognino et~al.}(2018{\natexlab{b}})}]{Bolognino:2018oth}
\bibinfo{author}{\bibfnamefont{A.~D.} \bibnamefont{Bolognino}} \bibnamefont{et~al.}, \bibinfo{journal}{Eur. Phys. J. C} \textbf{\bibinfo{volume}{78}}, \bibinfo{pages}{772} (\bibinfo{year}{2018}{\natexlab{b}}), \eprint{1808.05483}.

\bibitem[{\citenamefont{Bolognino et~al.}(2019{\natexlab{a}})}]{Bolognino:2019yqj}
\bibinfo{author}{\bibfnamefont{A.~D.} \bibnamefont{Bolognino}} \bibnamefont{et~al.}, \bibinfo{journal}{Acta Phys. Polon. Supp.} \textbf{\bibinfo{volume}{12}}, \bibinfo{pages}{773} (\bibinfo{year}{2019}{\natexlab{a}}), \eprint{1902.04511}.

\bibitem[{\citenamefont{Bolognino et~al.}(2019{\natexlab{b}})}]{Bolognino:2019cac}
\bibinfo{author}{\bibfnamefont{A.~D.} \bibnamefont{Bolognino}} \bibnamefont{et~al.}, \bibinfo{journal}{PoS} \textbf{\bibinfo{volume}{DIS2019}}, \bibinfo{pages}{049} (\bibinfo{year}{2019}{\natexlab{b}}), \eprint{1906.11800}.

\bibitem[{\citenamefont{Celiberto et~al.}(2020)\citenamefont{Celiberto, Ivanov,  Papa}}]{Celiberto:2020rxb}
\bibinfo{author}{\bibfnamefont{F.~G.} \bibnamefont{Celiberto}}, \bibinfo{author}{\bibfnamefont{D.~{\relax Yu}.} \bibnamefont{Ivanov}},  \bibinfo{author}{\bibfnamefont{A.}~\bibnamefont{Papa}}, \bibinfo{journal}{Phys. Rev. D} \textbf{\bibinfo{volume}{102}}, \bibinfo{pages}{094019} (\bibinfo{year}{2020}), \eprint{2008.10513}.

\bibitem[{\citenamefont{Celiberto}(2021{\natexlab{b}})}]{Celiberto:2020wpk}
\bibinfo{author}{\bibfnamefont{F.~G.} \bibnamefont{Celiberto}}, \bibinfo{journal}{Eur. Phys. J. C} \textbf{\bibinfo{volume}{81}}, \bibinfo{pages}{691} (\bibinfo{year}{2021}{\natexlab{b}}), \eprint{2008.07378}.

\bibitem[{\citenamefont{Celiberto}(2023{\natexlab{a}})}]{Celiberto:2022kxx}
\bibinfo{author}{\bibfnamefont{F.~G.} \bibnamefont{Celiberto}}, \bibinfo{journal}{Eur. Phys. J. C} \textbf{\bibinfo{volume}{83}}, \bibinfo{pages}{332} (\bibinfo{year}{2023}{\natexlab{a}}), \eprint{2208.14577}.

\bibitem[{\citenamefont{Celiberto et~al.}(2018{\natexlab{b}})}]{Celiberto:2017nyx}
\bibinfo{author}{\bibfnamefont{F.~G.} \bibnamefont{Celiberto}} \bibnamefont{et~al.}, \bibinfo{journal}{Phys. Lett. B} \textbf{\bibinfo{volume}{777}}, \bibinfo{pages}{141} (\bibinfo{year}{2018}{\natexlab{b}}), \eprint{1709.10032}.

\bibitem[{\citenamefont{Bolognino et~al.}(2019{\natexlab{c}})}]{Bolognino:2019yls}
\bibinfo{author}{\bibfnamefont{A.~D.} \bibnamefont{Bolognino}} \bibnamefont{et~al.}, \bibinfo{journal}{Eur. Phys. J. C} \textbf{\bibinfo{volume}{79}}, \bibinfo{pages}{939} (\bibinfo{year}{2019}{\natexlab{c}}), \eprint{1909.03068}.

\bibitem[{\citenamefont{Bolognino et~al.}(2019{\natexlab{d}})}]{Bolognino:2019ccd}
\bibinfo{author}{\bibfnamefont{A.~D.} \bibnamefont{Bolognino}} \bibnamefont{et~al.}, \bibinfo{journal}{PoS} \textbf{\bibinfo{volume}{DIS2019}}, \bibinfo{pages}{067} (\bibinfo{year}{2019}{\natexlab{d}}), \eprint{1906.05940}.

\bibitem[{\citenamefont{Adachi et~al.}(2022)}]{AlexanderAryshev:2022pkx}
\bibinfo{author}{\bibfnamefont{I.}~\bibnamefont{Adachi}} \bibnamefont{et~al.} (\bibinfo{collaboration}{ILC International Community}) (\bibinfo{year}{2022}), \eprint{2203.07622}.

\bibitem[{\citenamefont{Celiberto et~al.}(2021{\natexlab{b}})}]{Celiberto:2021dzy}
\bibinfo{author}{\bibfnamefont{F.~G.} \bibnamefont{Celiberto}} \bibnamefont{et~al.}, \bibinfo{journal}{Eur. Phys. J. C} \textbf{\bibinfo{volume}{81}}, \bibinfo{pages}{780} (\bibinfo{year}{2021}{\natexlab{b}}), \eprint{2105.06432}.

\bibitem[{\citenamefont{Celiberto et~al.}(2021{\natexlab{c}})}]{Celiberto:2021fdp}
\bibinfo{author}{\bibfnamefont{F.~G.} \bibnamefont{Celiberto}} \bibnamefont{et~al.}, \bibinfo{journal}{Phys. Rev. D} \textbf{\bibinfo{volume}{104}}, \bibinfo{pages}{114007} (\bibinfo{year}{2021}{\natexlab{c}}), \eprint{2109.11875}.

\bibitem[{\citenamefont{Celiberto et~al.}(2022{\natexlab{a}})}]{Celiberto:2022zdg}
\bibinfo{author}{\bibfnamefont{F.~G.} \bibnamefont{Celiberto}} \bibnamefont{et~al.}, \bibinfo{journal}{Phys. Rev. D} \textbf{\bibinfo{volume}{105}}, \bibinfo{pages}{114056} (\bibinfo{year}{2022}{\natexlab{a}}), \eprint{2205.13429}.

\bibitem[{\citenamefont{Celiberto}(2022{\natexlab{a}})}]{Celiberto:2022keu}
\bibinfo{author}{\bibfnamefont{F.~G.} \bibnamefont{Celiberto}}, \bibinfo{journal}{Phys. Lett. B} \textbf{\bibinfo{volume}{835}}, \bibinfo{pages}{137554} (\bibinfo{year}{2022}{\natexlab{a}}), \eprint{2206.09413}.

\bibitem[{\citenamefont{Celiberto}(2024{\natexlab{a}})}]{Celiberto:2024omj}
\bibinfo{author}{\bibfnamefont{F.~G.} \bibnamefont{Celiberto}}, \bibinfo{journal}{Eur. Phys. J. C} \textbf{\bibinfo{volume}{84}}, \bibinfo{pages}{384} (\bibinfo{year}{2024}{\natexlab{a}}), \eprint{2401.01410}.

\bibitem[{\citenamefont{Anchordoqui et~al.}(2022)}]{Anchordoqui:2021ghd}
\bibinfo{author}{\bibfnamefont{L.~A.} \bibnamefont{Anchordoqui}} \bibnamefont{et~al.}, \bibinfo{journal}{Phys. Rept.} \textbf{\bibinfo{volume}{968}}, \bibinfo{pages}{1} (\bibinfo{year}{2022}), \eprint{2109.10905}.

\bibitem[{\citenamefont{Feng et~al.}(2023)}]{Feng:2022inv}
\bibinfo{author}{\bibfnamefont{J.~L.} \bibnamefont{Feng}} \bibnamefont{et~al.}, \bibinfo{journal}{J. Phys. G} \textbf{\bibinfo{volume}{50}}, \bibinfo{pages}{030501} (\bibinfo{year}{2023}), \eprint{2203.05090}.

\bibitem[{\citenamefont{Celiberto}(2022{\natexlab{b}})}]{Celiberto:2022rfj}
\bibinfo{author}{\bibfnamefont{F.~G.} \bibnamefont{Celiberto}}, \bibinfo{journal}{Phys. Rev. D} \textbf{\bibinfo{volume}{105}}, \bibinfo{pages}{114008} (\bibinfo{year}{2022}{\natexlab{b}}), \eprint{2204.06497}.

\bibitem[{\citenamefont{Celiberto}(2024{\natexlab{b}})}]{Celiberto:2024swu}
\bibinfo{author}{\bibfnamefont{F.~G.} \bibnamefont{Celiberto}}, \bibinfo{journal}{Particles} \textbf{\bibinfo{volume}{7}}, \bibinfo{pages}{502} (\bibinfo{year}{2024}{\natexlab{b}}), \eprint{2405.09526}.

\bibitem[{\citenamefont{Boussarie et~al.}(2018)}]{Boussarie:2017oae}
\bibinfo{author}{\bibfnamefont{R.}~\bibnamefont{Boussarie}} \bibnamefont{et~al.}, \bibinfo{journal}{Phys. Rev. D} \textbf{\bibinfo{volume}{97}}, \bibinfo{pages}{014008} (\bibinfo{year}{2018}), \eprint{1709.01380}.

\bibitem[{\citenamefont{Chapon et~al.}(2022)}]{Chapon:2020heu}
\bibinfo{author}{\bibfnamefont{E.}~\bibnamefont{Chapon}} \bibnamefont{et~al.}, \bibinfo{journal}{Prog. Part. Nucl. Phys.} \textbf{\bibinfo{volume}{122}}, \bibinfo{pages}{103906} (\bibinfo{year}{2022}), \eprint{2012.14161}.

\bibitem[{\citenamefont{Celiberto\mbox{, M. Fucilla}}(2022)}]{Celiberto:2022dyf}
\bibinfo{author}{\bibfnamefont{F.~G.} \bibnamefont{Celiberto\mbox{, M. Fucilla}}}, \bibinfo{journal}{Eur. Phys. J. C} \textbf{\bibinfo{volume}{82}}, \bibinfo{pages}{929} (\bibinfo{year}{2022}), \eprint{2202.12227}.

\bibitem[{\citenamefont{Celiberto}(2023{\natexlab{b}})}]{Celiberto:2023fzz}
\bibinfo{author}{\bibfnamefont{F.~G.} \bibnamefont{Celiberto}}, \bibinfo{journal}{Universe} \textbf{\bibinfo{volume}{9}}, \bibinfo{pages}{324} (\bibinfo{year}{2023}{\natexlab{b}}), \eprint{2305.14295}.

\bibitem[{\citenamefont{Celiberto\mbox{, A. Papa}}(2024)}]{Celiberto:2023rzw}
\bibinfo{author}{\bibfnamefont{F.~G.} \bibnamefont{Celiberto\mbox{, A. Papa}}}, \bibinfo{journal}{Phys. Lett. B} \textbf{\bibinfo{volume}{848}}, \bibinfo{pages}{138406} (\bibinfo{year}{2024}), \eprint{2308.00809}.

\bibitem[{\citenamefont{Celiberto et~al.}(2024{\natexlab{a}})\citenamefont{Celiberto, Gatto,  Papa}}]{Celiberto:2024mab}
\bibinfo{author}{\bibfnamefont{F.~G.} \bibnamefont{Celiberto}}, \bibinfo{author}{\bibfnamefont{G.}~\bibnamefont{Gatto}},  \bibinfo{author}{\bibfnamefont{A.}~\bibnamefont{Papa}} (\bibinfo{year}{2024}{\natexlab{a}}), \eprint{2405.14773}.

\bibitem[{\citenamefont{Celiberto}(2024{\natexlab{c}})}]{Celiberto:2024mrq}
\bibinfo{author}{\bibfnamefont{F.~G.} \bibnamefont{Celiberto}}, \bibinfo{journal}{Symmetry} \textbf{\bibinfo{volume}{16}}, \bibinfo{pages}{550} (\bibinfo{year}{2024}{\natexlab{c}}), \eprint{2403.15639}.

\bibitem[{\citenamefont{Chen et~al.}(2015)}]{Chen:2014gva}
\bibinfo{author}{\bibfnamefont{X.}~\bibnamefont{Chen}} \bibnamefont{et~al.}, \bibinfo{journal}{Phys. Lett. B} \textbf{\bibinfo{volume}{740}}, \bibinfo{pages}{147} (\bibinfo{year}{2015}), \eprint{1408.5325}.

\bibitem[{\citenamefont{Boughezal et~al.}(2015)}]{Boughezal:2015dra}
\bibinfo{author}{\bibfnamefont{R.}~\bibnamefont{Boughezal}} \bibnamefont{et~al.}, \bibinfo{journal}{Phys. Rev. Lett.} \textbf{\bibinfo{volume}{115}}, \bibinfo{pages}{082003} (\bibinfo{year}{2015}), \eprint{1504.07922}.

\bibitem[{\citenamefont{Monni et~al.}(2020)\citenamefont{Monni, Rottoli,  Torrielli}}]{Monni:2019yyr}
\bibinfo{author}{\bibfnamefont{P.~F.} \bibnamefont{Monni}}, \bibinfo{author}{\bibfnamefont{L.}~\bibnamefont{Rottoli}},  \bibinfo{author}{\bibfnamefont{P.}~\bibnamefont{Torrielli}}, \bibinfo{journal}{Phys. Rev. Lett.} \textbf{\bibinfo{volume}{124}}, \bibinfo{pages}{252001} (\bibinfo{year}{2020}), \eprint{1909.04704}.

\bibitem[{\citenamefont{Celiberto et~al.}(2022{\natexlab{b}})}]{Celiberto:2021tky}
\bibinfo{author}{\bibfnamefont{F.~G.} \bibnamefont{Celiberto}} \bibnamefont{et~al.}, \bibinfo{journal}{PoS} \textbf{\bibinfo{volume}{EPS-HEP2021}}, \bibinfo{pages}{589} (\bibinfo{year}{2022}{\natexlab{b}}), \eprint{2110.09358}.

\bibitem[{\citenamefont{Celiberto et~al.}(2022{\natexlab{c}})}]{Celiberto:2021fjf}
\bibinfo{author}{\bibfnamefont{F.~G.} \bibnamefont{Celiberto}} \bibnamefont{et~al.}, \bibinfo{journal}{SciPost Phys. Proc.} \textbf{\bibinfo{volume}{8}}, \bibinfo{pages}{039} (\bibinfo{year}{2022}{\natexlab{c}}), \eprint{2107.13037}.

\bibitem[{\citenamefont{Celiberto et~al.}(2022{\natexlab{d}})}]{Celiberto:2021txb}
\bibinfo{author}{\bibfnamefont{F.~G.} \bibnamefont{Celiberto}} \bibnamefont{et~al.}, \bibinfo{journal}{PoS} \textbf{\bibinfo{volume}{PANIC2021}}, \bibinfo{pages}{352} (\bibinfo{year}{2022}{\natexlab{d}}), \eprint{2111.13090}.

\bibitem[{\citenamefont{Hamilton et~al.}(2013)}]{Hamilton:2012rf}
\bibinfo{author}{\bibfnamefont{K.}~\bibnamefont{Hamilton}} \bibnamefont{et~al.}, \bibinfo{journal}{JHEP} \textbf{\bibinfo{volume}{05}}, \bibinfo{pages}{082} (\bibinfo{year}{2013}), \eprint{1212.4504}.

\bibitem[{\citenamefont{Bagnaschi et~al.}(2023)}]{Bagnaschi:2023rbx}
\bibinfo{author}{\bibfnamefont{E.}~\bibnamefont{Bagnaschi}} \bibnamefont{et~al.}, \bibinfo{journal}{Eur. Phys. J. C} \textbf{\bibinfo{volume}{83}}, \bibinfo{pages}{1054} (\bibinfo{year}{2023}), \eprint{2309.10525}.

\bibitem[{\citenamefont{Banfi et~al.}(2024)}]{Banfi:2023mhz}
\bibinfo{author}{\bibfnamefont{A.}~\bibnamefont{Banfi}} \bibnamefont{et~al.}, \bibinfo{journal}{JHEP} \textbf{\bibinfo{volume}{02}}, \bibinfo{pages}{023} (\bibinfo{year}{2024}), \eprint{2309.02127}.

\bibitem[{\citenamefont{Celiberto et~al.}(2023)}]{Celiberto:2023uuk_article}
\bibinfo{author}{\bibfnamefont{F.~G.} \bibnamefont{Celiberto}} \bibnamefont{et~al.}, \bibinfo{journal}{\emph{Proceedings of Moriond QCD}}  (\bibinfo{year}{2023}), \eprint{2305.05052}.

\bibitem[{\citenamefont{Celiberto et~al.}(2024{\natexlab{b}})}]{Celiberto:2023eba_article}
\bibinfo{author}{\bibfnamefont{F.~G.} \bibnamefont{Celiberto}} \bibnamefont{et~al.}, \bibinfo{journal}{PoS} \textbf{\bibinfo{volume}{RADCOR2023}}, \bibinfo{pages}{069} (\bibinfo{year}{2024}{\natexlab{b}}), \eprint{2309.11573}.

\bibitem[{\citenamefont{Celiberto et~al.}(2024{\natexlab{c}})}]{Celiberto:2023nym}
\bibinfo{author}{\bibfnamefont{F.~G.} \bibnamefont{Celiberto}} \bibnamefont{et~al.}, \bibinfo{journal}{PoS} \textbf{\bibinfo{volume}{EPS-HEP2023}}, \bibinfo{pages}{390} (\bibinfo{year}{2024}{\natexlab{c}}), \eprint{2310.16967}.

\bibitem[{\citenamefont{Hentschinski et~al.}(2021)}]{Hentschinski:2020tbi}
\bibinfo{author}{\bibfnamefont{M.}~\bibnamefont{Hentschinski}} \bibnamefont{et~al.}, \bibinfo{journal}{Eur. Phys. J. C} \textbf{\bibinfo{volume}{81}}, \bibinfo{pages}{112} (\bibinfo{year}{2021}), \eprint{2011.03193}.

\bibitem[{\citenamefont{Celiberto et~al.}(2022{\natexlab{e}})}]{Celiberto:2022fgx}
\bibinfo{author}{\bibfnamefont{F.~G.} \bibnamefont{Celiberto}} \bibnamefont{et~al.}, \bibinfo{journal}{JHEP} \textbf{\bibinfo{volume}{08}}, \bibinfo{pages}{092} (\bibinfo{year}{2022}{\natexlab{e}}), \eprint{2205.02681}.

\bibitem[{\citenamefont{Nefedov}(2019)}]{Nefedov:2019mrg}
\bibinfo{author}{\bibfnamefont{M.~A.} \bibnamefont{Nefedov}}, \bibinfo{journal}{Nucl. Phys. B} \textbf{\bibinfo{volume}{946}}, \bibinfo{pages}{114715} (\bibinfo{year}{2019}), \eprint{1902.11030}.

\bibitem[{\citenamefont{Dawson et~al.}(2022)}]{Dawson:2022zbb}
\bibinfo{author}{\bibfnamefont{S.}~\bibnamefont{Dawson}} \bibnamefont{et~al.} (\bibinfo{year}{2022}), \eprint{2209.07510}.

\bibitem[{\citenamefont{Celiberto\mbox{, A. Papa}}(2023)}]{Celiberto:2023rtu}
\bibinfo{author}{\bibfnamefont{F.~G.} \bibnamefont{Celiberto\mbox{, A. Papa}}} (\bibinfo{year}{2023}), \eprint{2305.00962}.

\end{thebibliography}
\endgroup

\end{document}